# From psychological traits to safety warnings: three studies on recommendations in a smart home environment

Federica Cena[1,†], Cristina Gena[1,†], Claudio Mattutino[1,†], Michele Mioli[1,†] and Fabiana Vernero[1,*,†]

[1]*University of Turin, Department of Computer Science, C.so Svizzera 185, 10149 Turin, Italy*

**Abstract**

In this paper, we report on three experiments we have carried out in the context of the EMPATHY project, with the aim of helping users make better configuration choices in a smart home environment, and discuss our results. We found that there are psychological traits, such as Need for Cognition, which influence the way individuals tend to use recommendations, that there are non obvious relationships between the perceived usefulness of recommendations in different domains and individuals' ability to exploit suggestions on configuration choices, and that detailed, easy-to-understand security explanations are more persuasive than simple security warnings, when it comes to make decisions on the applicability of rules which might cause privacy and security risks.

## 1. Introduction

One of the main functions of recommender systems is to help users make better *preferential choices* [1] from a large set of items [2] by providing personalized and non-personalized [3] suggestions. For example, recommender systems are often employed to provide users with suggestions about products to buy, music to listen to, or movies to watch [4].

Recommender systems can support users' decision-making processes [2] and user preferential choices [1] by providing personalized and non-personalized [3] suggestions.

In the context of a smart environment, which integrates a plethora of heterogeneous, connected IoT components in a single system aimed at supporting its users in carrying out their tasks [5], recommender systems can be applied to make configuration tasks more manageable for end-users who often lack the technical knowledge needed to understand and, therefore, control their IoT environment [6]. Assuming that configuration tasks consist in the definition of trigger-action rules[1], end-users need to make choices (and, therefore, may benefit from receiving support) on different aspects, among which are: what smart objects to include in a smart environment; whether to define a rule to determine an interaction pattern between two given objects; what event should trigger a certain rule; and what action should be carried out when a trigger event occurs.

While recommendations *per se* can help to make configuration tasks easier in general, a couple of specific aspects are worth noticing. Firstly, the actual perception, acceptance and appreciation of recommendations can depend on individual features, such as end-users' personality, a factor which plays an important role in decision-making processes [7], or their familiarity with and inclination to exploit recommendations in other scenarios. Secondly, when defining trigger action rules, end-users may involuntarily create serious risks for their privacy and/or for the security of the smart environment itself [8]. Hence, a reliable trigger-action platform for an IoT environment should not only suggest possible rules but also alert end-users whenever the rules they are defining may determine privacy and security risks.

In this paper, we report on three experiments we have carried out in the context of the EMPATHY project[2], with the aim of helping users make better configuration choices in a smart home environment, and discuss our results. In the first experiment (Section 3.0.1) we investigated whether psychological traits (Locus of Control, Self-efficacy, and Need for Cognition) impact the perceived usefulness of recommendations. In the second experiment (Section 3.0.2) we studied the correlation between the perceived usefulness of recommendations in different domains and users' performances in carrying out configuration tasks with the help of recommendations. In the third experiment (Section 3.0.3), we investigate end-users' perception of warning messages which inform them of potential privacy and security risks. To complete the paper, Section 2 presents the prototype platform we have used as a basis for our experiments, and Section 4 discusses our results.

## 2. The prototype recommendation platform

In the context of the EMPATHY project, we have built a prototype platform[3] to experiment with recommendation services aimed at supporting end-users in the definition of trigger/action rules for a smart home environment. Such a platform offers the following functionality, among others:

- A series of **user modeling services** which exploit tests based on standard IPIP (International Personality Item Pool) scales[4] to offer external applications a way to model their users as far as the four related traits of Self-Efficacy[9], Need for Cognition[10], Locus of Control[11] and Mindset [12] are concerned [6].
- A **series of recommendation services** which suggest smart objects (either a *trigger* or an *action* object) to couple with an input object chosen by the end-user. Example rules are also suggested for each trigger-action association [13].[5]
- A **security/safety warning** service which provides information on the potential privacy and security risks of trigger/action rules. Such information is retrieved by querying an external service provided by colleagues at the University of Salerno and consists in labels which identify the relevant category of risk (*innocuous*, *personal*, *physical*, and *cybersecurity*), as well as in more extensive explanations (for more details on the identification of risk categories and the provisioning of safety/security information, see [8]).

---

[1] An example trigger-action rule, meant for a smart home context, could be: if a weather station (trigger object) measures that the level of humidity in the air is above a certain threshold (trigger event), the automatic irrigation system (action object) is disabled (action event)[6].

[2] http://www.empathy-project.eu

[3] https://app.empathy.di.unito.it

[4] https://ipip.ori.org/

## 3. The experiments

### 3.0.1. First experiment: the influence of psychological traits on the perceived usefulness of rule recommendations

The goal of our first experiment was to assess how helpful (in terms of performance) and useful *trigger*, *action*, and *rule recommendations* are to users who are given a (simulated) configuration task, and if certain personality traits might have an impact on the way in which participants would view recommendations.

In particular, we focused on the following constructs: Locus of Control, Self-efficacy, and Need for Cognition, since they suit the constructive tasks of EUD. In fact, Locus of Control [11] refers to people's beliefs about how much control they have over the events of their lives; Self-efficacy [9] can be defined as people's belief in their capacity to exercise control over their own functioning and over events that may affect their lives; finally, Need for Cognition [10] reflects the extent to which people are inclined towards demanding cognitive activities.

Sixty-four participants (68.4% in the 18-25 age range, 26.2% in the 26-35 age range, and 3.2% in the 36-44 age range) took part in our experiment, where they were asked to write three rules to determine the behavior of their smart homes: for each rule, they had to choose a trigger object and an action object, then they had to write the rule's logic in natural language. Participants in the experimental group could access our prototype platform (see Section 2) to receive rule recommendations[6]. Participants answered several questions about their user experience, while their performance in rule definition was assessed by three independent domain experts.

---

[5] Notice that knowledge on suitable trigger-action associations was derived by applying the *association rules* technique on an IFTTT dataset (https://ifttt.com/).

[6] An example rule recommendation if participants select "weather station" as the trigger object and "lamp" as the action object is: *Change color of Hue bulb to blue when rain is detected*. An example rule written by participants for the same trigger object, with "smart garden" as the action object, is *If the weather station detects rain and low temperatures, the smart garden system will only water plants located indoors.*

Our results seem to confirm our hypothesis that exposure to examples and recommendations of possible rules increases performance in a configuration task, in general. In addition, Need for Cognition influences the way individuals tend to use the recommendations when performing the task, and this has an impact on their perception of usefulness and performance. In fact, users with a high Need for Cognition are more likely to use recommendations, when these are available, and they may appreciate and benefit from the aid [6].

### 3.0.2. Second experiment: the influence of perceived recommendation usefulness on recommendation effectiveness

Similarly to our first experiment (Section 3.0.1), our second experiment was also aimed at assessing the effectiveness and perception of recommendations in the EUD domain, with a larger number of participants and slightly different tasks. One hundred and twenty-five participants, aged 18-24 (77%), 25-34 (17%), 35-55 (4%), >44(3%), 29% males and 71% females, were asked to perform simple activities contextualized in scenarios relevant to the Internet of Things, such as home automation, and to answer some questions about themselves and the rule suggestions (recommendations) they received. More specifically, since one of our goals was to understand whether perceived recommendation usefulness in domains different from EUD influences users' behaviour when dealing with typical EUD tasks (the rationale being that users might have cross-domain preferences for and abilities to deal with recommendations), we asked participants to assess the usefulness of recommendations received on e-commerce websites such as Amazon (where following a recommendation might imply a relatively strong personal investment) and on media providers such as Netflix (where following recommendations has a lower cost), respectively, by means of rating scales [14]. Similarly to our first experiment, participants were also asked to carry out simulated configuration tasks, consisting in the definition of possible home automation rules with given trigger/action objects, with the help of recommendations.

We carried out a correlational study using Pearson's Correlation Coefficient. We found a medium correlation between the perceived usefulness for the two kinds of recommendations (r= 0.36), meaning that users who appreciate recommendations regarding items to buy (Amazon) also appreciate recommendations about movies to watch (Netflix), but there are some differences. Interestingly, we also found that participants' evaluation of recommendations on Amazon (r=0.98) and Netflix (r=0.83) is related to their performance in carrying out the proposed online activities with the help of recommendations. On the contrary, the relationship between participants' evaluation of recommendations and their performance in a configuration task where they received no recommendations was weaker (Amazon: r=0.42; Netflix: r=0.33).

### 3.0.3. Third experiment: safety warnings

Our third experiment was a small-scale pilot study aimed at studying end-users' perception of warning messages which inform them of potential privacy and security risks.

Sixteen participants, equally divided among males and females, aged 18-74 years old, were randomly assigned to two groups, each of which was exposed to safety warnings provided in a specific format, either simple labels identifying the relevant category of risk, or more detailed explanations. More specifically, participants were asked to interact with a prototype platform for the configuration of a smart home environment, to browse a series of predefined trigger/action rules, and to state whether they would apply them or not, and why. After that, the researcher showed them the safety warnings, if any, that the prototype platform would display in case the rules were actually applied, and asked participants whether they would confirm their original choice. In addition, participants were invited to express any further comments. Participants also answered several questions regarding their user experience and the perceived usefulness of safety warnings.

Interestingly, our results show that all the participants who received detailed explanations judged them useful, while this was true only for 75% of the participants in the other group (who were provided only with labels). Coherently with this observation, most participants (62.5%) who received detailed explanations believed that they influenced their final decision (whether to apply the rule or not) and made security/safety problems clear, while only 37.5% of the participants who received short explanations, consisting only in risk category labels, shared this idea, with most participants being neutral or negative. It is worth noticing that user behavior during the experiment confirms participants' answers, therefore highlighting the fact that detailed explanations are more effective. In fact, participants who received detailed explanations revised their initial choice more often, especially in the case of rules where the potential risk was not obvious. On the contrary, short explanations were not enough to make previously unnoticed risks clear, so participants in this group very rarely changed their initial choice.

## 4. Conclusion

In this paper, we presented three empirical studies we carried out in the context of the EMPATHY project, with the aim of making configuration choices more manageable for non-expert end-users. Our results show that there are psychological traits, such as Need for Cognition, which influence the way individuals tend to use the recommendations when performing configuration tasks. The actual impact of such traits should be further studied, so as to be able to better adapt recommendations to the personality of end-users. We also found that there is some relationship between the perceived usefulness of recommendations in different domains and individuals' ability to exploit suggestions on configuration choices in the context of a constructive task. Finally, we observed that end-users are more persuaded by detailed security explanations, and do not find it obvious to simply trust a system which provides security warnings when necessary. All our results are preliminary and the here discussed topics will be further studied in future work by combining and interplaying of quantitative and qualitative evaluation methodologies for enhancing the comprehension of evaluation data, as already suggested by other evaluations in recommender systems context [15].

## Acknowledgments

This work is partially supported by the Italian Ministry of University and Research (MIUR) under grant PRIN 2017 "EMPATHY: EMpowering People in deAling with internet of THings ecosYstems".

The first and second experiment were carried out in collaboration with and with the help of Barbara Treccani and Massimo Zancanaro (University of Trento), whom we thank.

For the third experiment, we thank Bernardo Breve (University of Salerno) for his consultancy on security aspects.